\documentstyle[epsfig]{aipproc}

\begin{document}
\title{Phase Diagram of the Lattice Restricted Primitive Model}

\author{Ronald Dickman$^{\dagger,}$
\thanks{electronic address: dickman@fisica.ufmg.br}
\thanks{On leave of absence from: Department of Physics and Astronomy,
Herbert H. Lehman College, City University of New York,
Bronx, NY, 10468-1589.}
 and
George Stell$^{\ddagger}$\thanks{electronic address: gstell@sbchm1.sunysb.edu}}
\address{$^{\dagger}$Departmento de F\'\i sica,
Universidade Federal de Minas Gerais, Caixa Postal 702,
30161-970 Belo Horizonte - MG, Brasil  \\
$^{\ddagger}$Department of Chemistry,
State University of New York at Stony Brook,
Stony Brook, NY 11794-3400, U.S.A.   }

\maketitle

\begin{abstract}
We present a comprehensive study
of the lattice restricted primitive model, i.e.,
a lattice gas consisting of an equal number of positively and
negatively charged particles interacting via
on-site exclusion and a $1/r$ potential.
On the cubic lattice, Monte Carlo simulations show a line of N\'eel points
separating a disordered,
high-temperature phase from a phase with global antiferromagnetic
order. At low temperatures the (high-density) ordered phase 
coexists with the (low-density) disordered phase.  The N\'eel line meets the
coexistence curve at a tricritical point, $T_t \simeq 0.14$, 
$\rho_t \simeq 0.4$.   A simple mean-field analysis is in qualitative
agreement with simulations.
\end{abstract}

\section*{Introduction}

It has long been realized that the presence of an
appreciable concentration
of free ions in a system of overall charge neutrality gives rise
to a variety of thermodynamic features not found in un-ionized systems.
For example, in the restricted primitive model (RPM), a system of charged
hard-sphere anions and cations, all with equal charge magnitude $|q|$
and diameter $\sigma$, one has the famous limiting law established by
the work of Debye and H\"uckel \cite{dhxx} in 1923,
\begin{equation}
-\frac{A^{ex}}{k_BT V} \rightarrow \frac{\Gamma_D^3}{12 \pi} \;\;\;
\mbox{ as } \;\; \rho \rightarrow 0, \;\; T \;\; \mbox{fixed}.
\end{equation}
Here $A^{ex}$ is the Helmholtz free energy (excess to the 
ideal-gas condition), $V$ is the system volume,
$k_B$ Boltzmann's constant, $T$ the temperature,
$\rho$ the total number density of anions and cations, and $\Gamma_D$
is the inverse Debye length, $\Gamma_D^2 \sigma^2 = 4 \pi \rho q^2/k_BT$.
This is in marked contrast to an un-ionized fluid, for which $A^{ex}/k_BTV$
approaches a second virial-coefficient term proportional to $\rho^2$ rather
than $\rho^{3/2}$.

Results pertaining to phase transitions and criticality in the RPM did not
emerge until much after the work of Debye and H\"uckel.  It was not until
1976 that a systematic study by Stell, Wu, and Larsen appeared with the 
conclusion that the RPM should exhibit a liquid-gas coexistence curve 
with a critical point \cite{swl76}.  They further concluded that the 
critical density is much lower than for a simple argon-like fluid model, 
but found that that the shape and location of the coexistence curve was
very sensitive to the details of the approximations that must inevitably
be used in such a theoretical study.  Monte Carlo simulations have yielded
results fully consistent with their conclusions, but only in the last
few years have the studies of different groups
converged toward common values for the critical density and 
temperature of the RPM \cite{orkoulas,caillol,valleau}.

As one of us has discussed in earlier studies \cite{stell92,stell95},
the RPM can be thought of as a spin system with a long-ranged 
antiferromagnetic interaction

\begin{equation}
J(r) = -\frac{q^2}{r\epsilon},
\end{equation}
and one-dimensional spins ${\bf s}_i$ with $|{\bf s}_i| = 1$;

\begin{equation}
\phi_{ij}(r) = -J(r) {\bf s}_i \cdot {\bf s}_j \;.
\end{equation}
(Since the ``spins" are charges, we should perhaps refer to the
interaction as antiferroelectric rather than antiferromagnetic.)
The lattice-gas version of such a spin fluid is simply the spin-1
Ising model with $|{\bf s}| = 1$, 0, or -1, with ${\bf s} = 0$ representing
a vacant site.  We shall refer to this model, which forms the
subject of this report, as the
{\it lattice restricted primitive model} (LRPM).

For a lattice gas with a nearest-neighbor $J({\bf r})$ instead of a Coulombic
$J({\bf r})$, the spin-1 Ising model is a Blume-Capel model \cite{blume,capel},
which becomes equivalent, at full occupancy, to a nearest-neighbor
spin-$\onehalf$ Ising model that is exactly-soluble on many two-dimensional
lattices \cite{mccoy,baxter}.  On the square lattice one finds a N\'eel point
as one lowers the temperature in the absence of an external field.
Moreover, although exact results are lacking in three dimensions,
one expects for bipartite lattices such as the simple cubic, that as one
lowers the density along
the $\lambda$-line of N\'eel points in the $\rho$-$T$ plane, one encounters
a tricritical point below which the lattice gas phase-separates into
paramagnetic and antiferromagnetic phases.
This immediately raises the question whether this remains true for
a lattice gas with a Coulombic $J({\bf r})$.

Several years ago, in order to better understand the LRPM, we
initiated Monte Carlo simulations on a simple cubic lattice.
Our results are fully consistent with the presence of a $\lambda$-line of
N\'eel points terminating in a tricritical point, below which
(in $T$) two phases coexist.  Preliminary results were reported in
Ref. \cite{stell99}; here we provide more extensive results as well as
details of our simulation procedure.
We have supplemented our simulations with a mean-field analysis
that exploits the similarity between the LRPM and a
nearest-neighbor lattice gas.
H$\mbox{\o}$ye and Stell \cite{hs97} have also made a more general study of the spin-1
Ising model that included the solution of the mean-spherical approximation
(MSA) for a range-parametrized $J({\bf r})$.  As discussed by these authors,
the MSA is not appropriate to study criticality
and phase separation in the spin-1 model with Coulombic $J({\bf r})$
(although, as they show, it does yield the correct Debye-H\"uckel limiting law).
The H$\mbox{\o}$ye-Stell study, however, goes beyond the MSA with a number of general
observations that strongly support tricriticality for a Coulombic $J({\bf r})$.

More recent studies made by Ciach and Stell \cite{ciach} of ionic models that include
the continuum-space and lattice RPM reveal that generically such models can be
expected to manifest criticality, tricriticality, or both, depending upon the
precise forn of their Hamiltonians.  For example, the work suggests that some
extended-core lattice models may well show both liquid-gas like criticality and,
at higher densities, a $\lambda$-line with associated tricriticality.  For the
RPM, the critical properties that follow from application of RG analysis are
found to be in the Ising universality class.

\section*{Model}

We consider a lattice gas of particles interacting via site exclusion (multiple
occupancy forbidden) and a Coulomb interaction $u(r_{ij}) = s_i s_j /r_{ij}$,
where $r_{ij} = |{\bf r}_i - {\bf r}_j|$ is the distance separating the particles
(located at lattice sites ${\bf r}_i $ and ${\bf r}_j $), 
and $s_i = +1$ or $-1$
is the charge of particle $i$.   Exactly half the $N$ particles are positively charged;
the remainder are negative.  They are restricted to a simple cubic lattice of
$V = L^3$ sites, with periodic boundaries. 
We assume a lattice constant $a$ of unity,
and adopt units in which $q^2/a k_B = 1$,
$q$ being the magnitude of the charge.  In what follows we treat charge, length,
energy and temperature as dimensionless.

We note in passing that in lattice simulations of ionic systems an alternative
definition of the potential is possible, i.e., $u({\bf r}_{ij}) = s_i s_j  \Phi({\bf r}_{ij})$,
where $\Phi$ is the Green's function for Poisson's equation on the lattice in
question.  While the simple $1/r$ potential and the lattice Green's function
differ somewhat at short distances, they have the same asymptotic (large $r$)
behavior, and we expect (qualitatively) the same phase diagram in either case.

In this study we restrict attention to the simple cubic lattice.  This lattice, like the
body-centered cubic, admits a decomposition into two sublattices (the sites on
one sublattice having all nearest neighbors in the other sublattice), which
obviously facilitates formation of an ordered state resembling an ionic crystal.
Indeed, it was proven some time ago that the LRPM on the simple cubic lattice
exhibits long-range order at sufficiently low temperatures and high 
fugacities \cite{lieb80}.
The model presents, as we will show, a strong tendency
to assume a NaCl-like ordered state at low temperatures.  It would be interesting
to study the model on the face-centered cubic lattice or another structure that
frustrates antiferromagnetic ordering.

\section*{Mean-Field Analysis}

The energy of the lattice restricted primitive model is

\begin{equation}
\label{energy}
U = \frac{1}{2} \sum_{i,j} \frac{s_i s_j} {r_{ij}},
\end{equation}

\noindent where it is understood that the particles occupy distinct lattice sites.  
We consider the LRPM with independent variables $T$ (temperature) and 
$\rho \equiv N/V$.  
It is helpful to think of this system as a three-state (i.e., spin-1)
antiferromagnetic Ising model with long-range interactions ($J \sim 1/r$).
On a fully occupied lattice ($\rho = 1$) we characterize ordering by the sublattice charge disparity or staggered
magnetization $\phi$.  In the context of MFT, this allows us to replace Eq. (\ref{energy}) by
the corresponding expression for a {\em nearest-neighbor} (NN) lattice gas, multiplied by a suitable factor.  

Consider first the disordered system.  Since the neighbors of any given particle are equally likely
to bear the same or opposite charges, the mean-field estimate for the energy is zero, just as for
the NN system. Next, suppose there is sublattice ordering;  let $\rho_+^A $ be the fraction
of sites in sublattice $A$ occupied by positive particles, etc., and let

\begin{equation}
\rho_+^A = \frac {\rho}{2} (1+\phi) \; ; \;\;\;\;\;\;\;    \rho_+^B = \frac {\rho}{2} (1-\phi) 
\end{equation}
\begin{equation}
\rho_-^A = \frac {\rho}{2} (1-\phi) \; ; \;\;\;\;\;\;\;    \rho_-^B = \frac {\rho}{2} (1+\phi) ,
\end{equation}
so that

\begin{equation}
\phi = \frac {\rho_+^A - \rho_-^A}{\rho},
\end{equation}
and 
\begin{equation}
\rho_+^A + \rho_-^A = \rho_+^B + \rho_-^B = \rho.
\end{equation}

Using these, the mean-field estimate for the energy of a
pair of nearest-neighbor sites is
\begin{eqnarray}
\nonumber
u_{NN} &=& \rho_+^A \rho_+^B +  \rho_-^A \rho_-^B -  \rho_+^A \rho_-^B -  \rho_-^A \rho_+^B \\ 
         &  = & - \rho^2 \phi^2.
\end{eqnarray}
Similarly, the MF estimate for the energy of a pair of second-neighbor sites (which 
must belong to the same sublattice) is $ + \rho^2 \phi^2/\sqrt 2$, and so on.  Thus MFT
gives the energy of the LRPM as
\begin{equation}
U_{MF} = \frac {\rho^2 \phi^2}{2} \sum_{i,j} \frac{(-1)^S}{r_{ij}} ,
\end{equation}
where $S= +1 (-1)$ if sites $i$ and $j$ belong to the same (different) sublattices.  The sum
(including the factor of 1/2) is simply the electrostatic energy of an ionic crystal.  For the
simple cubic lattice, 

\begin{equation}
\frac{1}{2} \sum_{i,j} \frac{(-1)^S}{r_{ij}} = -3 V \frac{\alpha}{6} ,
\end{equation}

\noindent  as $V \rightarrow \infty$, where $\alpha = 1.7457$ is the Madelung constant
\cite{tosi}.  Thus the mean-field energy per site is $u_{MF} = - \alpha \rho^2 \phi^2/2$,
while the corresponding expression for a system with nearest-neighbor interactions
is $-3 \rho^2 \phi^2$.  In the context of MFT, then, we may replace the $1/r$ potential
with a nearest-neighbor interaction.

From the preceding analysis it appears that we may treat the LRPM, in MFT, as if it
were a nearest-neighbor lattice gas, but with an energy (and temperature) scale that is
smaller by a factor of $\alpha /6 \simeq 0.2913$.  A more meaningful way of relating the
temperature scales, however, is to compare the energy change attending an elementary
excitation, in this case, the interchange of a nearest-neighbor pair in a perfectly
ordered lattice.  In the simple cubic lattice this raises the energy by 
$\Delta U_{NN} = 20$ (the nearest-neighbor interaction $J=1$),
since ten nearest-neighbor
pairs have like charges after the exchange.  For the ionic crystal (NaCl structure) the
corresponding energy change is $4(\alpha -1) \simeq 2.9903$, so that 
$\Delta U_{LRPM} \simeq 0.1495 \Delta U_{NN}$.  Since the mean-field critical temperature
for the (fully occupied) nearest-neighbor lattice gas is 6 on the cubic lattice, the
corresponding result for the LRPM is about 0.9.  The best numerical estimate
for the lattice gas is $T_c = 4.5115$; the corresponding LRPM value is 0.674.
(Extrapolation of our simulation results to $\rho = 1$ yields $T_c \approx 0.6$.)

We may now develop the MFT of the LRPM by studying the nearest-neighbor
lattice gas, bearing in mind the difference in temperature scales explained above.  To estimate
the entropy as a function of $\rho$ and $\phi$, we note that the number of allowed configurations
on a sublattice of $V/2$ sites is $(V/2)!/[N_+! N_-! N_v!] $ where $N_+$ ($N_-$) is the number of positive 
(negative) particles and
$N_v = [(V/2) - N_+ -N_-]$ the number of vacant sites on the sublattice.  Using
$N_{\pm} = (1\pm \phi) \rho V/4$, etc., and Stirling's formula, we obtain the entropy per site,

\begin{equation}
s = -\rho \ln \rho - (1\!-\! \rho) \ln (1\!- \!\rho) + \rho \ln 2 
     - \frac {\rho}{2} [(1+\phi) \ln(1+\phi) + (1-\phi) \ln (1-\phi) ] \;.
\label{entmf}
\end{equation}

Let $f = u - Ts$ be the Helmholtz free energy per site.  To investigate the possibility of a
free energy minimum with nonzero sublattice ordering $\phi$, we note that

\begin{equation}
\left ( \frac{\partial f}{\partial \phi} \right) _{\rho} = 
      \rho \left[ -6 \rho \phi + \frac {T}{2} \ln \frac{1+\phi}{1-\phi} \right] \;.
\end{equation}

\noindent A free energy minimum therefore implies that
\begin{equation}
\frac {12 \rho \phi}{T} = \ln \frac{1+\phi}{1-\phi} \;.
\label{min}
\end{equation}

\noindent Expanding the r.h.s. as $2\phi + 2\phi^3/3 + \cdots$, we see that solutions
with nonzero $\phi$ exist for $T < T_c = 6$ and $ \rho > \rho_c = T/T_c$.  For
$\rho \stackrel > \sim \rho_c$,

\begin{equation}
x \simeq \pm \sqrt { 3 \left[ \frac{\rho}{\rho_c} - 1 \right]} \;.
\label{neel}
\end{equation}

\noindent Thus $T = 6\rho$ is a line of second-order phase transitions (a N\'eel or lambda line),
separating the high-temperature phase ($x=0$) from one with sublattice ordering.

Next we investigate the stability of the uniform-density state.  Defining $f_p = f/\rho$ as the free
energy per particle, we may write the chemical potential as
\begin{equation}
\mu = f_p + \frac{p}{\rho} \;,
\label{chempot}
\end{equation}
where the pressure $p$ is given by
\begin{equation}
p = \rho^2 \left(\frac{\partial f_p}{\partial \rho} \right) _T 
  = -T \ln (1-\rho) - 3 \rho^2 \phi^2 \;.
\label{press}
\end{equation}
(In evaluating the derivative we use Eq. (\ref{min}) to eliminate $(\partial \phi/\partial \rho)_T$.)
The pressure in the disordered phase is that of a simple lattice gas; but for
$T < T_c$, $(\partial p /\partial \rho)_T $ suffers a discontinuity at $\rho_c$.
Using Eq. (\ref{neel}) we have that for $\rho \stackrel > \sim \rho_c$,

\begin{equation}
\left( \frac{\partial p}{\partial \rho} \right) _T = \frac {T}{1-\rho} - 
                            9\rho \left(\frac{3\rho}{\rho_c} - 2 \right) \;,
\label{invcmp}
\end{equation}
so that
\begin{equation}
\left( \frac{\partial p}{\partial \rho} \right) _{T; \rho = \rho_c} = 
             T \left(\frac{1}{1-\rho_c} - \frac{3}{2} \right) \;.
\end{equation}

\begin{figure}[t!] 
\centerline{\epsfig{file=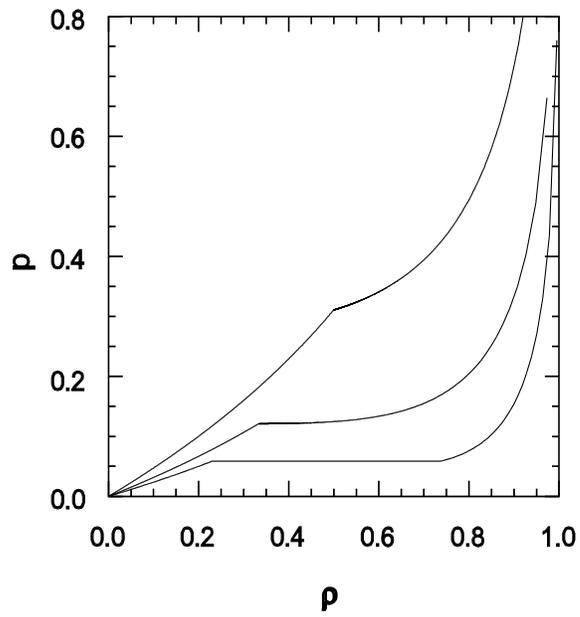,height=3.5in,width=3.5in}}
\vspace{2pt}
\caption{Mean-field theory predictions for the pressure versus
density at temperatures 0.449 (upper curve), $T_t =0.299$ (middle),
and 0.224 (lower).}
\label{dic1}
\end{figure}

\noindent Thus the inverse compressibility vanishes when $\rho_c = 1/3$, which
implies $T=2$.  This marks a tricritical point, at which the N\'eel line meets the
coexistence curve.  (For $T < 2$ the extension of the N\'eel line, $T = 6 \rho$,
is in fact a spinodal.)  Recalling the temperature rescaling, our MFT gives a
tricitical temperature of 0.3 for the LRPM, or $T_t = 0.225$ if we use the
best estimate ($T_c = 4.5115$) in place of the mean-field result ($T_c = 6$)
for the simple lattice gas.
Typical pressure-density curves are shown in Figure \ref{dic1}.

\begin{figure}[b!] 
\centerline{\epsfig{file=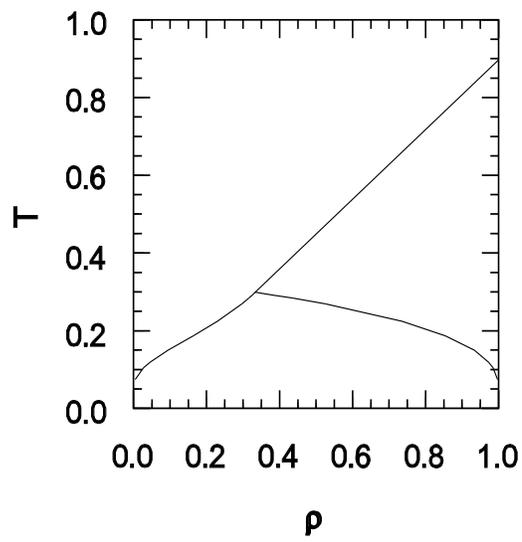,height=3.5in,width=3.5in}}
\vspace{2pt}
\caption{Mean-field prediction for the phase diagram of the LRPM.}
\label{dic2}
\end{figure}

We construct the coexistence curve by finding the densities $\rho_1$ and $\rho_2$
for which $p(\rho_1,T) = p(\rho_2,T) $ and $\mu(\rho_1,T) = \mu(\rho_2,T) $.
This is done numerically, through an iterative procedure. 
A first guess for $\rho_2$ is the mean of $\rho_s^+$
(the high-density spinodal, found by setting the r.h.s. of Eq.(\ref{invcmp}) to zero),
and $\rho_m $, the density at which the pressure equals $p(\rho_s^-)$,
its value on the low-density spinodal.  A first guess for $\rho_1$ is the density
at which $p(\rho_1) = p(\rho_m)$, given by the low-density isotherm,
$\rho = 1 - e^{-p/T}$. We then compare chemical potentials; if 
$\mu(\rho_2) > \mu(\rho_1)$, our guess for $\rho_2$ lies above the coexistence 
value, and {\em vice-versa}.  By refining the estimate for $\rho_2$, recalculating
$\rho_1$, and repeating the comparison, we rapidly converge on the coexisting
densities.  For temperatures $T \leq 0.5$ one may use 
$\phi = 1 - 2e^{-12 \rho_2/T} \simeq 1$ to show that $\rho_2 \simeq 1 - e^{-3/T}$ and
$\rho_1 \simeq 2 e^{-3 \rho_2/T}$.  The mean-field phase diagram for the LRPM
(with temperature scaled such that $T_t = 0.3$) is shown in Figure \ref{dic2}.

\section*{Simulation Methods}

\subsection*{Electrostatic energy of a periodic array}

This subsection presents the multipole expansion
as an efficient alternative to Ewald summation for evaluating the
electrostatic energy of a periodic array of point charges.
Our approach is closely related to the one employed by Ladd
in simulations of ionic and dipolar fluids \cite{ladd}.
We require the
electrostatic energy of an infinite, periodic array of cells containing
point charges.  Each cell harbors $N/2$ positive charges and
and equal number of negative charges; the position of particle
$i$ in cell $j$ is

\begin{equation}
{\bf x}_{i,j} = {\bf x}_i + {\bf R}_j \;,
\end{equation}
where {\bf x}$_i$ is the position in the central cell, and

\begin{equation}
{\bf R}_j = a (n_x \hat{\bf x}  +n_y \hat{\bf y} +n_z \hat{\bf z}),
\end{equation}
where $n_{\alpha}$ may assume any integer value.  (The cell structure
is that of a simple cubic lattice; the positions {\bf x}$_i$
may or may not be restricted to a lattice.)

We write the electrostatic energy of the array as
$U = (1/2)\sum_{i =1}^{N} q_{ i} \Phi_{i}$, where
$\Phi_{\rm i} $ is the potential at the position of charge i.
When calculating $\Phi_{i} $ we consider
a cube of side $a$ (oriented parallel to the periodic array), centered
on charge $i$, and write 

\begin{equation}
\Phi_{i} = \Phi_{i, 0} + \Phi_{i, a},
\label{decomp}
\end{equation}
where $\Phi_{i, 0}$ is the contribution from the charges (other than i)
within the cube, and $\Phi_{i, a}$ is the potential at the center of
the cube (i.e., the position of charge i) due to the infinite collection
of periodic image cells.  (These cells include the image of charge $i$,
as they must, if they are to be neutral.)
The lattice sum defining $\Phi_{i,a}$ is conditionally convergent, and
is often treated using the Ewald technique \cite{deleeuw}.
Here, however, we evaluate $\Phi_{ i, a}$ via a multipole
expansion. The contribution to $\Phi_{i, a}$ due to cell $j$
($j$ is a shorthand for the cell indices $n_x$, $n_y$, and $n_z$), is

\begin{equation}
\Phi_{i, j} = 4 \pi \sum_{l,m} \frac{q_{lm}}{2l + 1}
\frac{Y_{lm}(\theta_j,\phi_j)}{R_j^{l+1}}.
\label{mpe}
\end{equation}

\noindent Here $Y_{lm}$ is the spherical harmonic,
$R_j= a \sqrt{n_x^2 +n_y^2 + n_z^2}$,  $\cos \theta_j = an_z/R_j$,
$\tan \phi_j = n_y/n_x$,
and 
\begin{equation}
q_{l,m} =\sum_{ i = 1}^{N} q_i r_i^{l}
Y_{lm}^{*} (\theta_i, \phi_i),
\label{mumo}
\end{equation}

\noindent is the $(l,m)$ multipole moment of (any) cell,
with $(r_i,\theta_i,\phi_i)$ denoting the
position of particle $i$ relative to a convenient origin,
for example the cell center.  (The monopole moment $q_{00}$
vanishes by charge-neutrality.)

Summing the contributions from all cells (except the central one)
yields

\begin{equation}
\Phi_{i, a} = 4 \pi \sum_{l,m} \frac{q_{lm}}{2l + 1}
\sum_j \frac{Y_{lm}(\theta_j,\phi_j)}{R_j^{l+1}},
\label{mpep}
\end{equation}
where the sum on $j$ stands for a sum on $n_x$, $n_y$, and $n_z$ (not all zero).  This
sum over cells is independent of the charge distribution (which appears
only in the $q_{lm}$), and so need only be evaluated once. Many terms
vanish by symmetry.  Thus all terms with $l$ odd are zero
since $Y_{lm}$ is then an odd function of $\cos \theta$, while the
distribution of cells in the cubic lattice is symmetric about
$\theta = \pi$.  Symmetry also implies that nonzero contributions have
$m = 4n$ ($n$ integer), since for each cell with azimuthal angle $\phi$, there
are corresponding cells with angles $\phi + \pi/2$, $\phi + \pi$, and
$\phi + 3\pi/2$, and $1 + e^{im\pi/2} + e^{im\pi} + e^{3im\pi/2} = 4$
if $m = 4n$, and zero if $m$ is an integer not divisible by 4.
(Angle $\phi$ is not defined at the poles, but $Y_{lm}$ vanishes there
anyway for $m \neq 0$.)

Next we put the lattice sums that multiply the multipole moments
into a convenient form.  Using the relations

\begin{equation}
Y_{lm} (\theta,\phi) =  \sqrt{\frac{2l+1}{4\pi}\frac{(l-m)!}{(l+m)!} } \; 
P_l^m (\cos \theta) \; e^{im\phi},
\label{rel1}
\end{equation}

\begin{equation}
P_l^{-m} (x) =  (-1)^m \frac{(l-m)!}{(l+m)!}  \; P_l^m (x),
\label{rel2}
\end{equation}
and
\begin{equation}
Y_{l,-m} (\theta,\phi) =  (-1)^mY_{lm} (\theta,\phi),
\label{rel3}
\end{equation}
we can write Eq. (\ref{mpep}) in the form,

\begin{equation}
\Phi_{i, a} = \sum_{l \;\; even} \left\{ \overline{q}_{l0} \sum_j \frac{P_l (\cos \theta_j)}{R_j^{l+1}}
+ \sum_{m=4}^{l'} \frac{(l-m)!}{(l+m)!} 
\left[ \overline{q}_{lm} \sum_j \frac{P_l^m (\cos \theta_j) \; e^{im\phi_j}}{R_j^{l+1}} + 
({\rm c.c.}) \right] \right\},
\label{mpepa}
\end{equation}
where the sum on $m$ is restricted to integer multiples of 4, with $l' $ the largest such
multiple $\leq l$,  and ``c.c." denotes the complex conjugate of the preceding term.
The rescaled multipole moments are

\begin{equation}
\overline{q}_{l,m} =\sum_{i = 1}^{N} q_i r_i^{l}
P_l^m(\cos \theta_i) e^{-im \phi_i}.
\label{mumb}
\end{equation}

Cubic lattice symmetry implies that the lattice sums are real, since the azimuthal angles 
associated with cells at a given $R$ and $\theta$ are either
($0, \pi/2, \pi, 3\pi/2$), or ($\pi/4, 3\pi/4, 5\pi/4, 7\pi/4$), 
or take the form ($\pm \phi, \pi/2 \pm \phi, \pi \pm \phi, 3\pi/2 \pm \phi$).  Thus we may
write

\begin{equation}
\Phi_{i, a} = \sum_{l \;\; even} \sum_{m=4}^{l'} \overline{q}_{lm} S_{lm},
\label{mpepb}
\end{equation}
where $l'$ is the largest multiple of 4 that is $\leq l$, and
 
\begin{equation}
S_{l0} = \sum_j \frac{P_l (\cos \theta_j)}{R_j^{l+1}},
\label{slo}
\end{equation}
and for $m > 0$,

\begin{equation}
S_{lm} = 2\frac{(l-m)!}{(l+m)!}  \sum_j \frac{P_l^m(\cos \theta_j) \cos m\phi_j }{R_j^{l+1}},
\label{sloa}
\end{equation}
and the rescaled moments now take the form

\begin{equation}
\overline{q}_{l,m} =\sum_{i = 1}^{N} q_i r_i^{l}
P_l^m(\cos \theta_i) \cos m \phi_i.
\label{mumc}
\end{equation}
If $\overline{S}_{lm} $ is the sum for the {\it integer} lattice then
$S_{lm} = \overline{S}_{lm}/a^{l+1}$.  

The expansion actually begins at $l=4$.  To see this,
let $\{n_x,n_y,n_z\}_{R}$ be the set of
sites at distance $R$ from the origin of the (integer) cubic lattice.  Then

\begin{equation}
\overline{S}_{20} = \sum_j \frac{P_2 (\cos \theta_j)}{R_j^3} 
= \sum_R \frac{1}{R^3} 
\sum_{\{n_x,n_y,n_z\}_{R}} \frac{1}{2} (3 \cos^2 \theta -1).
\label{s2o}
\end{equation}
The inner sum is
\begin{equation}
\frac{1}{2R^2} \sum_{\{n_x,n_y,n_z\}_{R}} (3 n_z^2 - n_x^2 -n_y^2 - n_z^2) = 0,
\label{s2oi}
\end{equation}
since the cubic lattice is symmetric under permutations of $x$, $y$, and $z$.

Applied to the NaCl lattice (positive and negative charges on the two
sublattices), the expansion to order $l=10$ yields an energy per particle of
$-\alpha q^2/4 \pi \epsilon_0 a$ with $\alpha = 1.747561$.
(The standard value is 1.747565 \cite{tosi}.)
For this highly symmetric
arrangement the $l=4$ and 8 moments all vanish; the $l=10$ contribution
to the energy has a magnitude of about 0.03 that of the $l=6$ term.
The lattice sums $S_{lm}$ converge rather quickly; 
the following results were obtained by summing
over all sites with $R \leq 30$:

\[
\overline{S}_{40} = 3.108219   \;\;\;\;\;\;\;\;\;\;\;\;  
\overline{S}_{44} = 0.0370026
\]
\[
\overline{S}_{60} = 0.5733293   \;\;\;\;\;\;\;\;\;\;  
\overline{S}_{64} = -0.001592581
\]
\[
\overline{S}_{80} = 3.25929309   \;\;\;\;\;\;\;\;\;\;  
\overline{S}_{84} = 5.487025 \times 10^{-4}   \;\;\;\;\;\;\;\;\; 
\overline{S}_{88} = 0.035665665
\]
\[
\overline{S}_{10\;0} = 1.00922399   \;\;\;\;\;\;\;\;  
\overline{S}_{10\;4} = -1.84839558 \times 10^{-4}   \;\;\;\;\;\;\;
\overline{S}_{10\;8} = -4.2786935 \times 10^{-8}   .
\]

In the simulation routine we use these results to construct a lookup table giving the
total contibution of a pair of charges at sites {\bf r} and {\bf r}' to the potential energy.
The table entry includes the direct (minimum-image) contribution, the lattice-sum
contribution obtained via multipole expansion, and a dipolar contribution to be explained in the following subsection. 



\subsection*{Dipole Correction Term}

In this very brief subsection we call attention to an absolutely crucial 
point in the simulation of systems (on or off-lattice) with slowly decaying 
forces.  It is that the
energy of the fully periodic array {\it differs from} the infinite 
lattice sum (taken, e.g.,
in spherical order, as above), by a term proportional to the 
square of the cell dipole moment:
\begin{equation}
U = \frac{1}{2} \sum_i s_i \Phi_i 
  =  \frac{1}{2} \sum_i s_i [\Phi_{i, 0} + \Phi_{i, a}]
       + \frac{2 \pi}{3L^3} M^2
\label{dip1}
\end{equation}
where
\begin{equation}
{\bf M} = \sum_i s_i {\bf r}_i
\label{dipole}
\end{equation}
is the dipole moment of (any) cell.  This result was derived and discussed in
detail by de Leeuw, Perram, and Smith in 1980 \cite{deleeuw}.
It follows from a careful analysis
of the lattice sum (summed in a particular order, using a convergence factor) applies
to both the multipole respresentation of $\Phi_{\rm i, a}$ developed above, and to the
more familiar Ewald expansion.  We therefore add the 
term $2 \pi s_i s_j r_{ij}^2/3L^2$
to the contribution of charges $i$ and $j$ to the potential energy.
Some preliminary studies, which did not include this dipole term, yielded bizarre results,
such as the tendency for a system with $\rho \simeq 1/4$ to crystallize into a uniform-density body-centered cubic structure at low temperatures.

\subsection*{Canonical Ensemble Simulations}  

The majority of the data reported here were obtained in canonical ensemble simulations
following the usual Metropolis prescription, i.e., trial configurations that lower the energy
are accepted with probability 1, while those attended by an energy increase $\Delta U > 0$
are accepted with probability $e^{-\Delta U/T}$.  An initial configuration
is generated by placing $N$ particles onto the lattice at random, with double 
occupancy forbidden.  This naturally yields a rather high-energy configuration that must
be allowed to relax before the system attains equilibrium.  We generally used a configuration
relaxed at some temperature $T$ as the initial configuration for a study at a nearby temperature
$T'$.  The energy and order parameter (defined below) were monitored in order to check for
relaxation.

Trial configurations are generated by two kinds of ``moves."  The first is a 
single-particle move: a particle, chosen at random, is displaced by
$\Delta{\bf r} =  n_x \hat{\bf x} + n_y \hat{\bf y} + n_z \hat{\bf z}$, where 
$n_x$, $n_y$
and $n_z$ are independent, integer-valued random variables distributed uniformly
on the set $\{-m,...,m\}$.  (We generally used $m=2$.)  
If the trial site is occupied, the move is rejected; otherwise the energy change is evaluated and
the move accepted according to the Metropolis scheme. 
At low temperatures, each particle
typically has one or more neighbors of the opposite charge; single-particle moves, which
tend to disrupt these dipolar pairs, will have a low acceptance rate.  We therefore 
alternated single-particle moves with pair moves.  In this case we select a particle $i$ at random,
and check if one of the nearest-neighbor sites (also selected at random) harbors a particle.
(Call it $j$; if the site is vacant the trial ends.)
Given an occupied neighbor, particle $i$ is displaced by $\Delta${\bf r} as above, and particle
$j$ is placed at a randomly chosen nearest-neighbor site of particle $i$.  (Naturally, both trial
sites must be vacant for the move to occur.)  The move is accepted or rejected, as before,
based on the Metropolis criterion.  The results reported here represent averages over
(typically) 2 to 10 of runs, each consisting of on the order of $2 \times 10^4$ lattice
updates (i.e., moves per particle).  A run of this sort requires 1 - 2 days of cpu time on a fast DEC alpha machine.

In addition to the energy and the specific heat (obtained from the variance of the energy),
we monitored the charge charge-charge correlation function,
\begin{equation}
g_{qq} (r) = {\cal A} \sum_{|{\bf x}|=r} \langle s(0) s({\bf x}) \rangle, 
\label{gqq}
\end{equation}
and the density-density correlation function,
\begin{equation}
g_{\rho \rho} (r) = {\cal A} \sum_{|{\bf x}|=r} 
                     \langle |s(0)| |s({\bf x})| \rangle.
\label{grr}
\end{equation}

Here $s({\bf x}) $ = 1 (-1) if site {\bf x} is occupied by a positively
(negatively) charged particle, and is zero otherwise.  The normalization
${\cal A}$ is chosen such that $g_{\rho \rho} \rightarrow 1$ for 
$r \rightarrow \infty$.
We also studied the sublattice order parameter, $\phi$.  Writing
{\bf x} = $i \hat{\bf x} + j \hat{\bf y} + k \hat{\bf z}$, 
we have
\begin{equation}
\phi = \frac {1}{N} \sum_{i,j,k} \langle (-1)^{i+j+k} s(i,j,k) \rangle. 
\label{phi}
\end{equation}
The order parameter is unity for a fully ordered (i.e.,
NaCl structure) lattice, and zero when the sublattices
bear no net charge.

In order to gauge the onset of phase separation, we divided the system
into cells of $4 \times 4 \times 4$ sites, and studied cell-occupancy
histograms, $P(\rho)$.
In an homogeneous system we
expect $P(\rho)$ to be unimodal, while
a bimodal distribution indicates phase coexistence, and a broad or
plateau-like distribution incipient phase separation.

\subsection*{Grand Canonical Ensemble Simulations}  

In an effort to obtain independent, and, one hopes, more reliable information
on the coexistence curve, we performed simulations in the grand canonical
ensemble.  The procedure is complicated somewhat by the fact that we are
obliged to maintain charge neutrality, so that a pair of particles
(one positive, one negative) must be added or removed simultaneously.
Let $\lambda = e^{\beta \mu_P} $ be the pair fugacity, $\mu_P $ being the
associated chemical potential.  Then the grand partition function is
\begin{equation}
\Xi(\beta, \lambda, V) = \sum_{N \mbox{{\small even}}} \lambda^{N/2} \sum_{\{{\cal C}\}_N} 
e^{-\beta U({\cal C})} \;,
\label{gpf}
\end{equation}
where the second sum is over the set of nonoverlapping configurations of $N/2$ positive
and $N/2$ negative particles on a lattice of $V$ sites.

Let ${\cal C} $ be a valid configuration for $N$ particles, and ${\cal C}'$ one for $N+2$
particles, obtained by inserting particles (one positive, one negative) at some pair of vacant sites in ${\cal C}$.  (The insertion sites need
not be nearest neighbors.)
Let  $w({\cal C} \! \rightarrow \! {\cal C}') $ be the transition rate, in a Monte Carlo simulation,
for going from ${\cal C} $ to ${\cal C}' $ (pair insertion), 
and $w({\cal C}' \! \rightarrow \! {\cal C}) $ the rate for the reverse process (pair deletion).
The detailed-balance condition,
\begin{equation}
\frac {w({\cal C} \! \rightarrow \! {\cal C}') }{w({\cal C}' \! \rightarrow \! {\cal C})}
 = \lambda e^{- \beta [U({\cal C}') - U({\cal C})] } \;,
\label{detbal}
\end{equation}
may be realized in various ways.  In our simulations, pair addition and deletion steps are
as follows.  A fraction $g/2$ of the moves are addition attempts, and equal number are
deletion attempts; the remainder are single-particle and pair displacements (no change in $N$) as described in the preceding subsection.  (In practice we used $g=0.9$.)  In an
insertion attempt we choose two sites (anywhere in the system) at random.  If either or both
are occupied, the attempt fails; otherwise we
tentatively place a positively charged particle at one site, and a negative particle at the
other.  Given the starting configuration ${\cal C}$, the probability of realizing a particular
trial configuration ${\cal C}'$ is $g/2V^2$.  The new configuration is accepted with
probability
\begin{equation}
P_{ins} = \min \left[ 1, \frac{4 \lambda V^2 e^{- \beta [U({\cal C}') - U({\cal C})] }}
{(N+2)^2} \right] \;.
\label{pins}
\end{equation}
In a deletion attempt, we choose one of the positively charged particles at random, and
similarly one of the negative particles, and tentatively delete them.  Starting with configuration
${\cal C}'$ (with $N+2$ particles), the probability of realizing ${\cal C}$ as the trial configuration
is $2g/(N+2)^2$.  In this case we accept the new configuration with probability
\begin{equation}
P_{del} = \min \left[ 1, \frac{(N+2)^2 e^{- \beta [U({\cal C}) - U({\cal C}')] }}
{4 \lambda V^2} \right]  = \min [1, P_{ins}^{-1}] \;.
\label{pdel}
\end{equation}
The transition rates $w({\cal C} \! \rightarrow \! {\cal C}') = (g/2V^2) P_{ins}$
and $w({\cal C}' \! \rightarrow \! {\cal C}) = [2g/(N+2)^2] P_{del}$ are
readily seen to satisfy detailed balance.

\section*{Results}

In this section we summarize our simulation results for
the phase diagram of the LRPM.  We
studied the energy, specific heat and order parameter
as functions of temperature, for a series of 12 or so
density values, ranging from about $\rho = 0.02$ to 0.85.
The data presented below were obtained in canonical ensemble simulations
with system size $L=16$ unless otherwise noted.

\begin{figure}[b!] 
\centerline{\epsfig{file=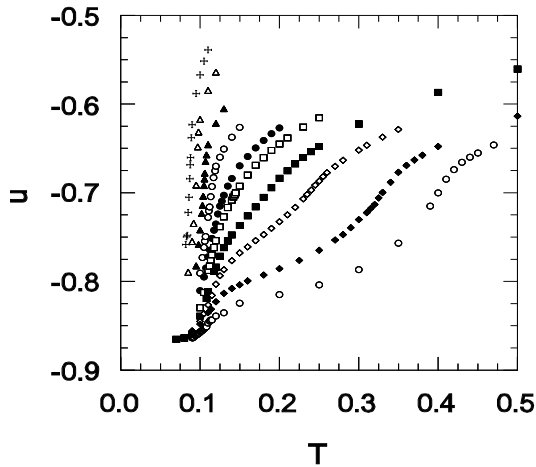,height=3.1in,width=3.4in}}
\vspace{2pt}
\caption[]{Energy per particle versus temperature.
Densities (left to right) $\rho = 0.04$,
0.08, 0.16, 0.25, 0.35, 0.4, 0.5, 0.6, 0.75 and 0.85.}
\label{dic3}
\end{figure}

\begin{figure}[t!] 
\centerline{\epsfig{file=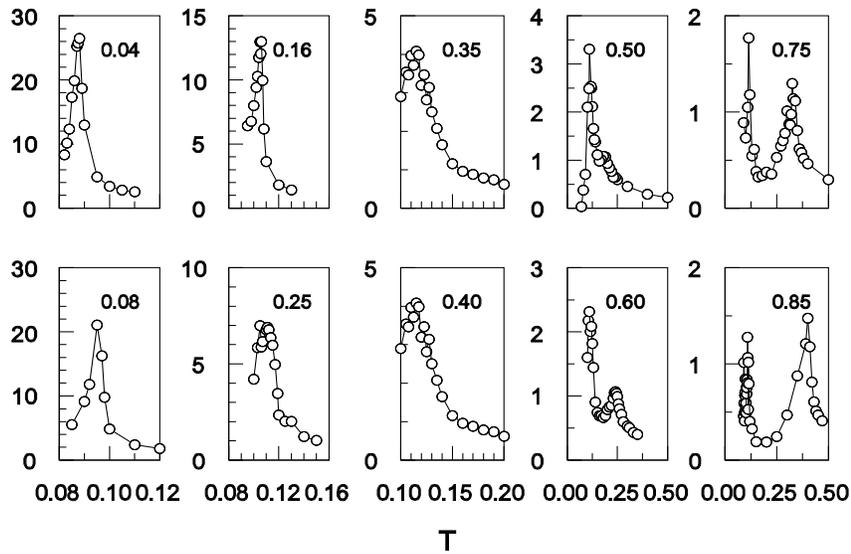,height=3.1in,width=3.4in}}
\vspace{2pt}
\caption{Specific heat versus
temperature for densities as indicated.}
\label{dic4}
\end{figure}

\begin{figure}[b!] 
\centerline{\epsfig{file=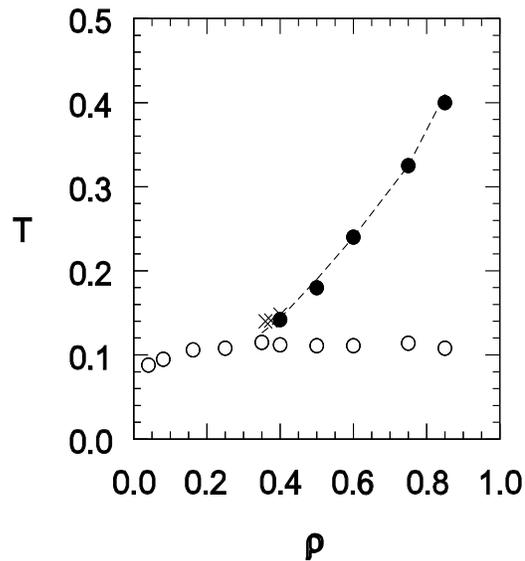,height=3.5in,width=3.5in}}
\vspace{2pt}
\caption[]{Loci of specific heat maxima (open and filled circles:
$L=16$; crosses: $L=20$). The dashed line denotes the position of the
maximum of $|d \phi/dT|$.}
\label{dic5}
\end{figure}

\begin{figure}[t!] 
\centerline{\epsfig{file=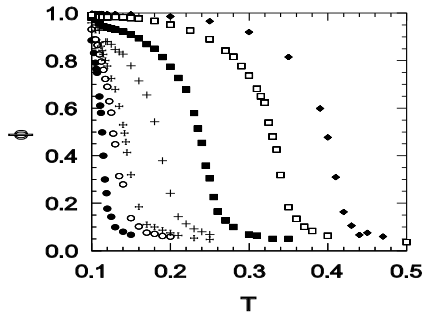,height=3.1in,width=3.3in}}
\vspace{2pt}
\caption[]{Order parameter versus temperature for densities (left to right)
0.25, 0.35, 0.4, 0.5, 0.6, 0.75, and 0.85.}
\label{dic6}
\end{figure}

Figure \ref{dic3} shows the energy and Figure \ref{dic4} the
specific heat versus temperature for a number of different densities.
All the specific heat curves show a peak at a temperature 
$T_1$ that increases with density for small densities, and then seems to saturate at
around $T = 0.11$ for $\rho \geq 0.35$.  The height of this peak 
decreases rapidly with increasing density.  For densities $\geq 0.4$,
there is a second peak, at a temperature $T_N > T_1$, which increases roughly
linearly with $\rho$.  
The locations of the specific heat peaks in the $\rho$-$T$
plane are shown in Figure \ref{dic5}; these loci appear to delineate a Neel 
line and a coexistence curve.  We examine their interpretation
in what follows.

\begin{figure}[b!] 
\centerline{\epsfig{file=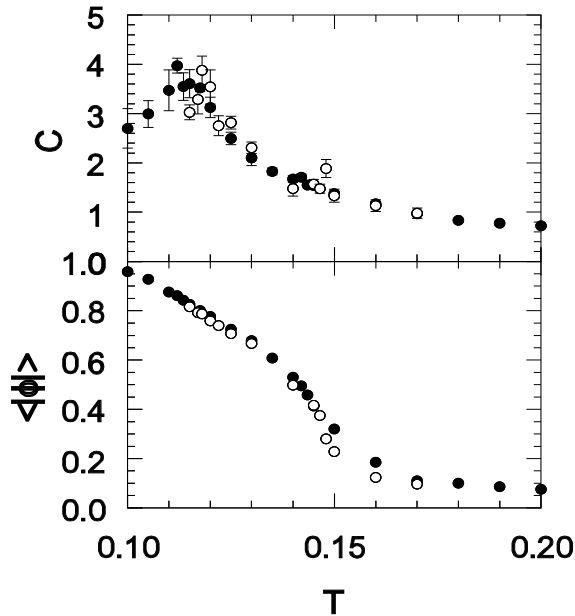,height=3.5in,width=3.5in}}
\vspace{2pt}
\caption[]{Upper graph: specific heat,
$\rho = 0.4$.  Filled circles: 
$L=16$; open circles:
$L=20$.  Lower graph: order parameter for $\rho = 0.4$; symbols 
as in upper graph.}
\label{dic7}
\end{figure}

It is tempting to associate the high-temperature specific heat peak with 
the onset of global sublattice ordering.  Studies of the order parameter and of the
charge-charge correlation function support this interpretation.  In 
Figure \ref{dic6}
we plot the mean value of $|\phi|$ versus temperature for various densities.
In the thermodynamic limit, this quantity should be strictly zero in the
disordered phase, and should increase continuously
as $T$ is reduced below $T_N$.
Here, of course, we must anticipate finite-size rounding of the critical
singularity, if any exists.  The data nevertheless give strong support for
$T_N$ marking an order-disorder transition.
In fact the temperature at which $|d\phi/dT|$ takes its maximum value
falls very close to that of the specific heat maximum, as 
shown in Figure \ref{dic5}.

Further support is provided by the comparison between two system
sizes (for the same density, $\rho = 0.4$)
shown in Figure \ref{dic7}: the larger system presents a sharper 
variation in $|\phi|$ in the vicinty of the apparent transition.  The
high-$T$ specific heat peak, barely visible in the $L=16$ data, is 
quite prominent for $L=20$; both peaks shift to slightly higher temperatures 
for $L=20$.  All of this is consistent with the system exhibiting phase
transitions at $T_1$ and $T_N$.  We shall therefore refer to the set
of transitions commencing around $\rho = 0.4$ as the N\'eel line.

\begin{figure}[t!] 
\centerline{\epsfig{file=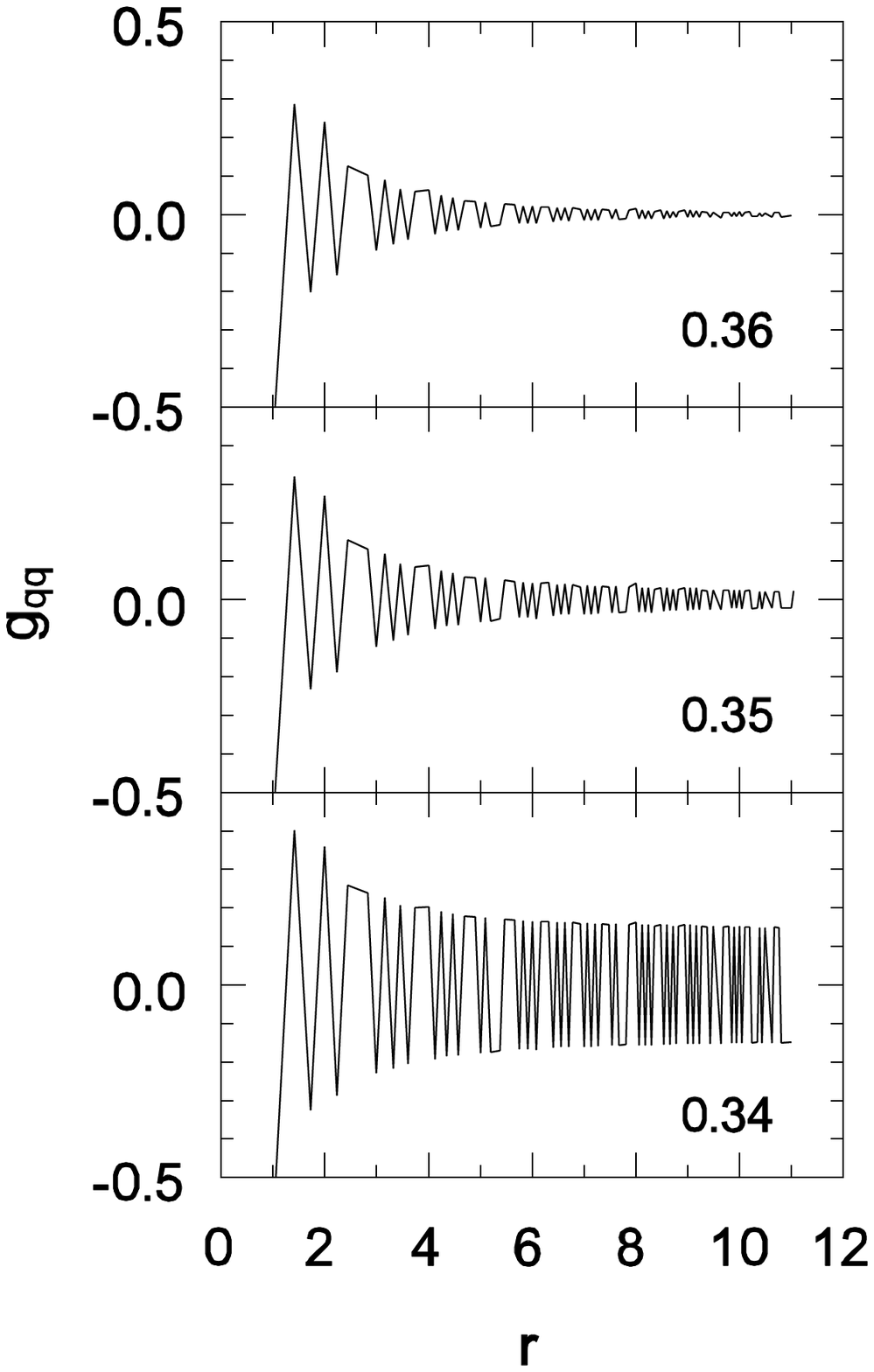,height=3.8in,width=3.3in}}
\vspace{2pt}
\caption[]{Charge-charge radial distribution function,
$\rho = 0.75$,
for three temperatures as indicated.}
\label{dic8}
\end{figure}

\begin{figure}[b!] 
\centerline{\epsfig{file=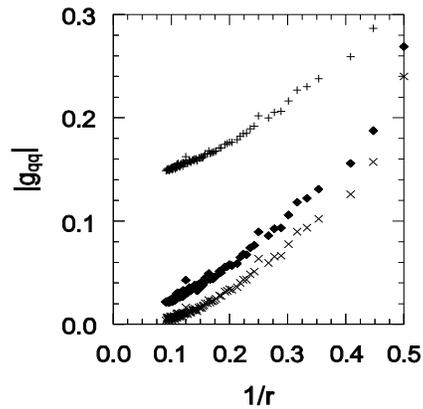,height=2.1in,width=2.25in}}
\vspace{2pt}
\caption[]{The data of Figure \ref{dic8} plotted versus $1/r$.  Lower set:
$T=0.36$; middle: $T=0.35$; upper: $T=0.34$.}
\label{dic9}
\end{figure}

\begin{figure}[t!] 
\centerline{\epsfig{file=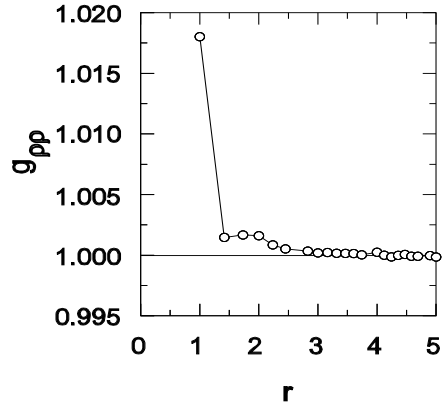,height=2.5in,width=3.0in}}
\vspace{2pt}
\caption[]{Density-density radial distribution function,
$\rho = 0.75$, $T=0.34$.}
\label{dic10}
\end{figure}

\begin{figure}[b!] 
\centerline{\epsfig{file=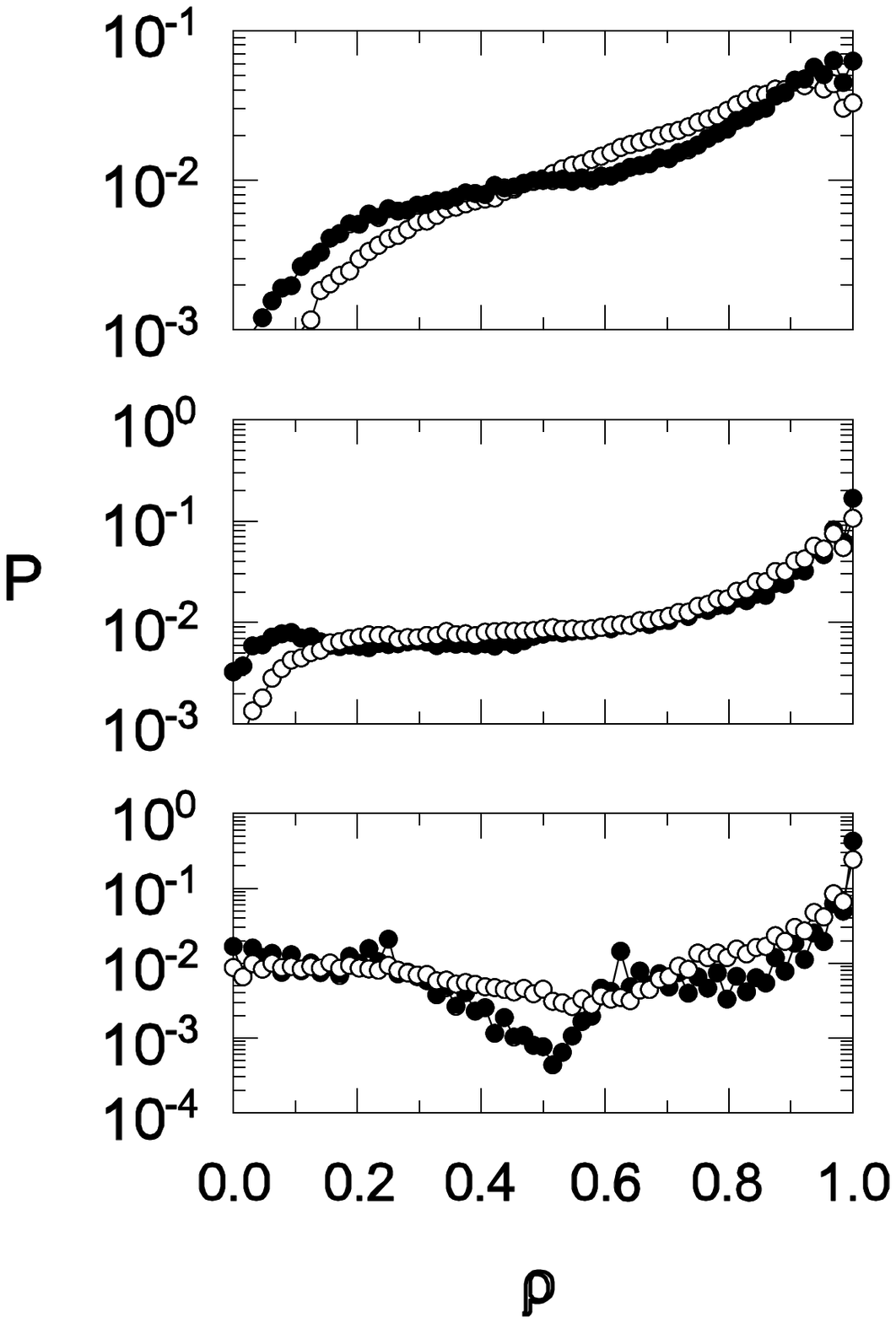,height=3.5in,width=2.8in}}
\vspace{2pt}
\caption[]{Cell-occupancy probabilities, $\rho = 0.75$.
Upper graph: open circles: $T=0.12$, filled circles $T=0.115$;
middle: open, $T=0.110$, filled, $T=0.105$; lower: open, $T=0.10$,
filled, $T=0.09$.}
\label{dic11}
\end{figure}

Further insight into the phase transition at or near the high-$T$ 
specific heat peak 
is afforded by the charge-charge correlation function $g_{qq} (r)$.
Figure \ref{dic8} shows $g_{qq} (r)$ for three 
temperatures in the neighborhood of $T_N$, for $\rho = 0.75$.  
In all of the cases shown, $g_{qq}$ exhibits damped oscillations.
At the highest
temperature ($T=0.36$) the oscillations evidently decay to zero; 
for $T=0.35$ the decay appears to be slower, while for $T=0.34$
they seem to persist.  A plot of $|g_{qq}|$ versus $1/r$ 
(Figure \ref{dic9}) 
yields a clearer idea of the large-$r$ behavior; from this graph it
is evident that $g_{qq}$ decays to zero for $T \geq 0.35$ but not for
$T=0.34$.  (The specific heat peaks at $T=0.325$.)
Thus the condition $lim_{r \rightarrow \infty} |g_{qq} (r)| > 0$
is a criterion for the ordered phase, and the
limiting value of $|g_{qq}|$ provides,
in principle, an alternative definition of the order parameter.
In practice, we are able to obtain $\phi$ with greater precision
(smaller statistical uncertainty), and so retain it as a measure
of global order.

\begin{figure}[t!] 
\centerline{\epsfig{file=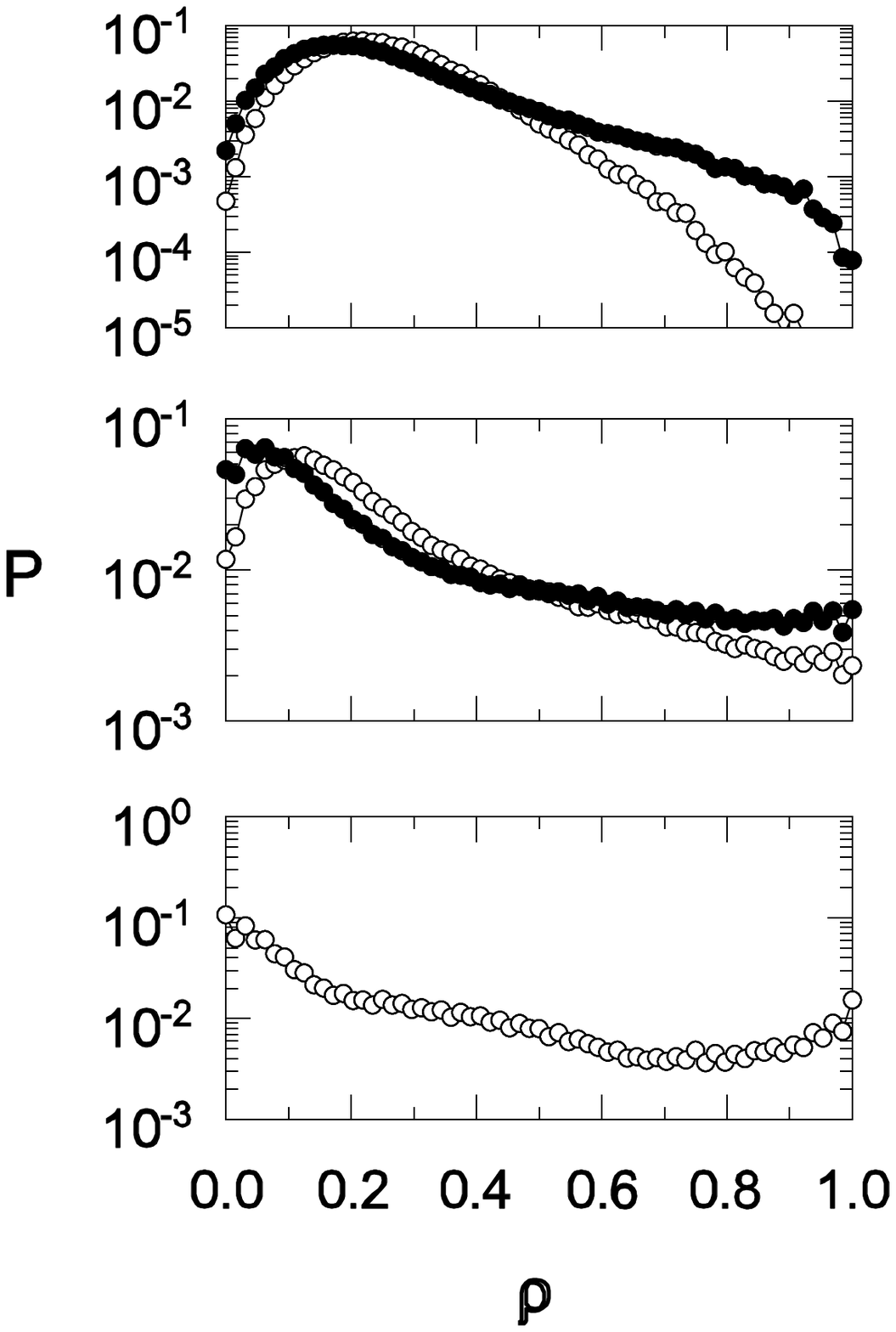,height=3.5in,width=2.8in}}
\vspace{2pt}
\caption[]{Cell-occupancy probabilities, $\rho = 0.25$.
Upper graph: open circles: $T=0.12$, filled circles $T=0.115$;
middle: open, $T=0.110$, filled, $T=0.105$; lower graph: $T=0.10$.}
\label{dic12}
\end{figure}

\begin{figure}[b!] 
\centerline{\epsfig{file=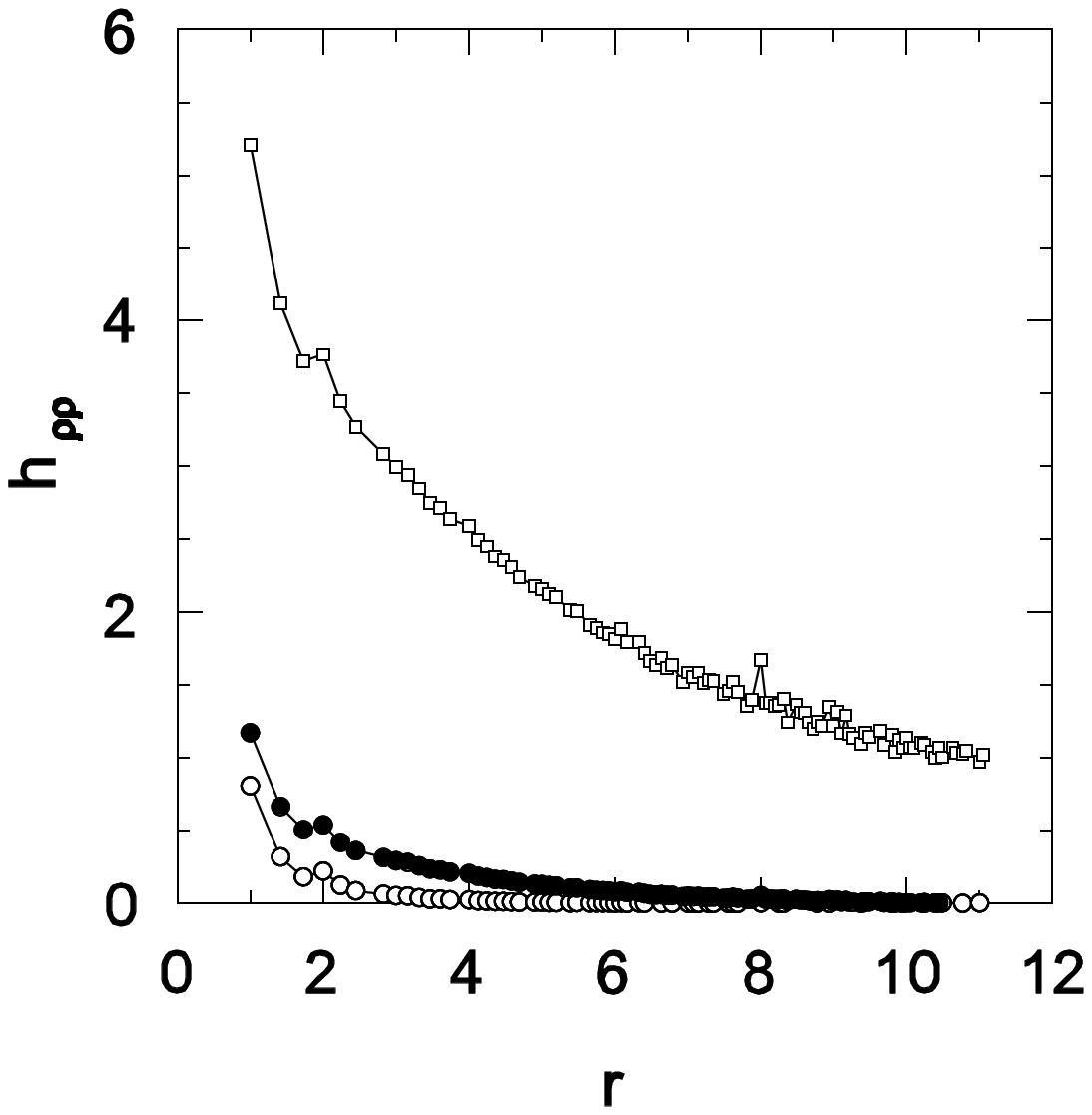,height=2.0in,width=2.0in}}
\vspace{2pt}
\caption[]{Density-density correlation function, $\rho = 0.25$,
$T=0.12$ (lower set), 0.115 (middle),
and 0.11 (upper).}
\label{dic13}
\end{figure}

It is worth noting that in the neighborhood of the lambda line, 
the density-density correlation function $g_{\rho \rho}$
shows little structure: it has a low-amplitude peak at unit distance
and then decays rapidly to its asymptotic value (see Figure \ref{dic10}).

We now turn to the line of low-temperature specific heat peaks.  
The mean-field theory results described above lead us
to interpret this as a coexistence curve, so that $T_1 (\rho)$ marks
a transition from a homogeneous system to coexisting high- and low-density
phases, the former disordered, the latter with sublattice ordering.
To better understand what is happening at $T_1$, we examine the occupancy
histograms.    
Some typical results for $\rho = 0.75$ are shown in Figure \ref{dic11}.
For $T=0.12$
the distribution is unimodal, and peaked near $\rho = 0.9$.  As the
temperature is lowered, the distribution broadens and the
peak moves to $\rho = 1$ and increases in amplitude, signaling the
emergence of a high-density phase.  For $T=0.11$ a plateau has appeared
for $0.2 \leq \rho \leq 0.7$, while for $T \leq 0.10$ the distribution
is bimodal.  (Note also the odd-even oscillations in $P$, especially
prominent for $T=0.09$.  These presumably represent a tendency to
local charge neutrality.)  All through the transition region, the
order parameter $\phi$ is very nearly unity; for this density, 
the transition near
$T_1$ is not associated with the onset of sublattice order.  (Such
order is already present in the uniform phase for $T_1 < T < T_N$.)

Figure \ref{dic12} shows a series of occupancy histograms for a much lower 
density, $\rho = 0.25$.
At the highest temperature shown, $T=0.12$, the distribution
peaks in the vicinity of $\rho = 0.2$ and decays rapidly for higher
densities.  For $T=0.115$ the distribution is already considerably broader,
and by $T=0.110$ a plateau near $\rho = 1$ has appeared.  At $T=0.105$
the distribution has become very broad, the main peak has shifted toward
$\rho = 0$, and a secondary peak has appeared at $\rho = 1$, indicating
separation into high- and low-density phases.  By $T=0.10$ the
transformation is complete: the histogram peaks at $\rho = 0$ and 
$\rho = 1$.  

It is interesting to observe the changes in the
density-density correlation function 
$h_{\rho \rho} (r) \equiv g_{\rho \rho} (r) -1$, in the neighborhood
of the transition.  In Figure \ref{dic13} we see that $h$ decays rapidly for
$T=0.12$ (the correlation length $\xi_{\rho} \simeq 0.9$), and is
somewhat longer-ranged ($\xi_{\rho} \simeq 2.2$) for $T=0.115$.  For
$T=0.11$ the decay is much slower ($\xi_{\rho} \simeq 6$).  At
temperatures below $0.11$, the system has phase-separated and 
we are no longer able to infer the large-$r$ limiting value of
$g_{\rho \rho} (r)$.  The histogram and correlation function data 
suggest that for $\rho = 0.25$, phase separation occurs at $T=0.105 - 0.110$,
in agreement with transition temperature of 0.108 indicated by
the specific heat maximum.  In contrast to what is seen at higher densities,
for $\rho = 0.25$ the transition at $T_1$ is accompanied by the
onset of sublattice order: $|\phi|$ is small for $T \geq 0.12$ and grows
rapidly in the range $T=0.105$ --- 0.115.  This is just what one
would expect, since the transition in this case marks the appearance
of a high-density phase, capable of supporting NaCl-like order.
(Sublattice ordering is associated with the transition at $T_1$
for $\rho \leq 0.25$.)

\begin{figure}[t!] 
\centerline{\epsfig{file=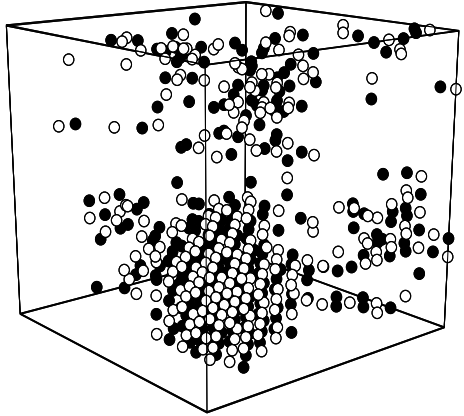,height=2.8in,width=3.4in}}
\vspace{2pt}
\caption[]{A typical configuration, $\rho=0.12$, $T=0.1$.
Filled and empty symbols distinguish the charges.}
\label{dic14}
\end{figure}

\begin{figure}[b!] 
\centerline{\epsfig{file=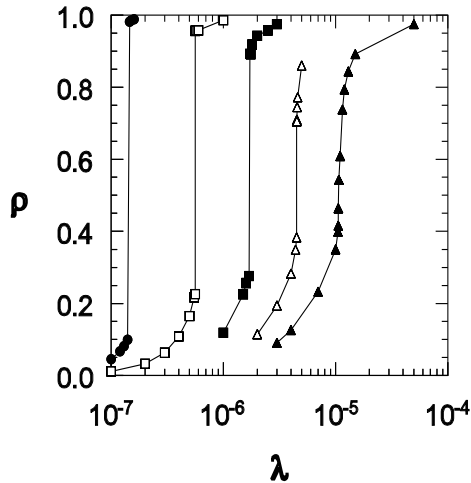,height=3.0in,width=2.8in}}
\vspace{2pt}
\caption[]{Density versus pair activity in
in grand canonical ensemble simulations ($L=16$) for temperatures
(left to right) $T=0.11$, 0.12, 0.13, 0.14 and 0.15.}
\label{dic15}
\end{figure}

Figure \ref{dic14} is a snapshot of a typical configuration of
the phase-separated system, in which we see a dense, crystal-like
agglomerate coexisting with a dilute vapor, comprised largely of dipolar
associations.

Using the specific heat data, together with results for the order
parameter and the charge-charge correlation function, we have
derived a series of estimates for points along the lambda line,
$T_N (\rho)$, for $\rho = 0.375$ --- 0.85.
The scatter amongst the values suggested by the
specific heat, order parameter, and correlation function data
yields an uncertainty estimate for $T_N (\rho)$.
Similarly, we use the
specific heat, cell occupancy histogram, and density-density
correlation function data to estimate the phase-separation
temperature $T_1 (\rho)$.

In an effort to determine the coexisting densities with greater
precision, we also performed grand canonical ensemble
(GCE) simulations.  Figure \ref{dic15} shows the dependence of the
density on the pair activity $\lambda$.  
For $T = 0.15$ the dependence is smooth, but for $T \leq 0.14$ there
is a discontinuity at $\lambda_c (T)$. Hysteresis effects
are minimal for $T \geq 0.12$, and 
the coexisting densities are given by the limiting values as we 
approach $\lambda_c $ from above and below.  The GCE
studies furnish an independent prediction
for the coexistence curve in the
range $T = 0.12$ --- 0.14.

\begin{figure}[t!] 
\centerline{\epsfig{file=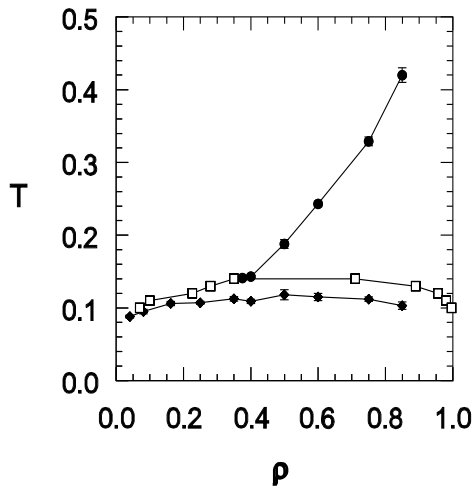,height=2.8in,width=2.8in}}
\vspace{2pt}
\caption[]{Best estimates for the location of the lambda line and
coexistence curve.  Filled symbols: canonical ensemble simulations;
open symbols: GCE simulations.}
\label{dic16}
\end{figure}

In Figure \ref{dic16} we plot our best estimates
for $T_N$ (from canonical simulations), and for the coexistence curve,
from both the canonical and GCE studies.
Evidently, the coexistence curve is shifted upward in the GCE 
simulations, whose results place the tricritical point
(i.e., the intersection of the lambda line and the coexistence curve),
at around $T=0.14$, as opposed to about $T=0.115$ in the canonical
studies.  Since the two ensembles are not required to give identical
results for finite systems, there is no inconsistency implied.
The grand canonical results, however, seem preferable on several
grounds.  First, the shape of the coexistence curve looks more reasonable
(more like the familiar Ising/lattice gas coexistence curve).  More
significantly, perhaps, the GCE curve actually meets the lambda line,
which appears to veer horizontally near $\rho = 0.4$, and shows no
sign of ever encountering the canonical coexistence curve.
Given the large surface effects implicit in phase coexistence in a 
small system, the GCE, which studies homogeneous phases, seems more
likely to give reliable results in a finite size simulation.
One may speculate that the large surface energy
depresses the transition temperature in the canonical ensemble simulations.
A more definitive answer to this question must await the results of
larger scale simulations.

\section*{Discussion and Summary}

We have determined the phase diagram of the lattice restricted primitive model
in extensive Monte Carlo simulations.
Our simulations reproduce the
principal features expected on the basis of general arguments and mean-field
theory, that is, a $\lambda$-line of N\'eel points meeting a coexistence curve
at a tricritcal point.
Our best estimate for the location of the tricritical point is
$T_t \simeq 0.14$, $\rho_t \simeq 0.4$, while mean-field theory 
predicts $T_t = 0.299$ and $\rho_t = 1/3$.  This comparison is
consistent with the tendency of mean-field theories to
overestimate transition temperatures, and underestimate transition
densities (i.e., to overestimate the stability of ordered phases).

During the completion of this work we learned of
Panagiotopoulos and Kumar's simulation results
for the LRPM \cite{panag}, which include a study of the effects of
varying the size of the hard core.  Their results for the case of
site-exclusion studied here support our conclusions
for the phase diagram, with the exception that the tricritical density
is somewhat larger ($\rho_t \simeq 0.48$ in their work) and the N\'eel
line appears to be nearly vertical in the vicinity of the tricritical
point.

Because our results come almost entirely from studies of a single system
size ($L=16$), it is difficult to assign meaningful uncertainties to
$T_t$, $\rho_t$, or, indeed, the location of the $\lambda$-line in the
$\rho$-$T$ plane.  The limited comparison we made with a system of
size $L=20$ (Figure \ref{dic7}) suggests that the true transition 
temperature (for the $\lambda$-transition) may be 5---10\% higher
than that observed for $L=16$.  We have also noted a sizeable
discrepancy ($\approx 25$\%) between canonical and grand canonical
simulations regarding the position of the coexistence curve along the
$T$-axis.  Studying larger systems is therefore the first order of business
for future simulations of the LRPM.  Since simulations are quite slow,
with many trial moves rejected at the low temperatures of interest,
even for the relatively small systems studied here, we anticipate the
need for non-Metropolis simulation methods utilizing histograms, 
multi-canonical sampling, and/or cluster dynamics to augment efficiency.

If such methods can be developed, a number of issues can
be explored.  It is of interest to understand the nature of
the transitions occuring along the $\lambda$-line and at the tricritical point,
to assign these their proper universality classes.
We may also anticipate that the interface between the dense and dilute
phases undergoes a roughening transition at a certain temperature.
Another topic worthy of investigation is the global phase diagram
as one goes from the Coulombic system studied here to a lattice gas with
nearest-neighbor interactions, via a family of models described by a
potential (e.g., of Yukawa form) having a range parameter.  The LRPM
with a lattice Green's function interaction, and on lattices that do not 
support antiferromagnetic order, are, as noted above, further topics for
future work.

\section*{Acknowledgments}

We are very grateful to Edgar Smith for correspondence
relating to the use of Eq. \ref{dip1}, which proved
invaluable for simulating the model properly.
We also thank Thanos Panagiotopoulos for helpful correspondence
and for communicating his simulation results prior to their publication.
R.D.acknowledges the support of the National Science Foundation for
his work done at Stony Brook.  G.S. acknowledges support by the Division
of Chemical Sciences, Office of Science, Office of Energy Research, U.S.
Department of Energy.

\end{document}